\documentclass[twocolumn,superscriptaddress, floatfix, amsfonts, aps]{revtex4-1}  
\usepackage{amsmath}
\usepackage{bm}
\usepackage{graphicx}
\usepackage{subfigure}
\usepackage{float}
\usepackage{mathrsfs}
\usepackage{footmisc}
\usepackage{multirow}
\usepackage{graphicx}
\usepackage{booktabs, ctable}
\usepackage{soul,xcolor}
\usepackage{xcolor, ulem, color, framed}
\usepackage{amssymb}
\usepackage[colorlinks, linkcolor=blue,anchorcolor=green,citecolor=red]{hyperref}

\graphicspath{{./figs/}}

\begin{document}

 \title{Bottomonium transport in p-Pb and Pb-Pb collisions at the LHC}

 \author{Baoyi Chen}
 \email{baoyi.chen@tju.edu.cn}

  \author{Bo Tong}

\affiliation{Department of Physics, Tianjin University, Tianjin 300354, China}

\begin{abstract}
We employ the Boltzmann transport model to study the sequential 
suppression pattern of $\Upsilon(1S,2S,3S)$ states 
in both small (p-Pb) and the large (Pb-Pb) collision systems at 
$\sqrt{s_{NN}}=5.02$ TeV. The cold nuclear matter effects happen before the formation of bottomonium, which is the same for different bottomonium states $\Upsilon(1S,2S,3S)$. The sequential suppression pattern of bottomonium states is treated as a signal of the hot medium effects, where bottomonium states suffer different magnitudes of the color screening and parton inelastic scatterings due to their different geometry sizes and binding energies. 
Bottomonium excited states are more easily dissociated with smaller binding energy via the final-state interactions. Including both cold and hot medium effects, transport model consistently explains the experimental data about bottomonium in both small and large collision systems.

\end{abstract}
\date{\today}

 \maketitle
 
%=======================================
\section{Introduction}

A new deconfined matter, which consists of elementary particles such as quarks and gluons~\cite{Bazavov:2011nk}, is believed to be created in relativistic heavy-ion collisions. The extremely hot medium is called ``Quark-Gluon Plasma'' (QGP). It has been extensively studied via aspects of heavy and light partons~\cite{Andronic:2015wma,Song:2010mg,Braun-Munzinger:2015hba,He:2014cla,Qin:2015srf,Chen:2019qzx,Grandchamp:2003uw,Yan:2006ve}. Due to the large mass, heavy quark and quarkonium production can be easily calculated via perturbative QCD. More than thirty years ago, heavy quarkonium was first proposed as a probe of the QGP by Matsui and Satz~\cite{Matsui:1986dk}. When the medium temperature is sufficiently high, heavy quark potential is screened by the thermal light partons in the QGP which reduces the binding energy of quarkonium and results in the melt of the bound states~\cite{Satz:2005hx}. Meanwhile, random collisions from thermal partons including gluon-dissociation~\cite{Zhu:2004nw} and parton quasi-free scatterings~\cite{Du:2017qkv} can also dissociate heavy quarkonium. 
Take bottomonium as an example, in heavy-ion collisions at Relativistic Heavy-Ion Collider (RHIC) and the Large Hadron Collider (LHC), hot medium effects give significant suppression on the production of bottomonium, which is characterized by the nuclear modification factor $R_{AA}$, defined as the ratio of bottomonium final production in nucleus-nucleus (AA) collisions and the product $N_{pp}^{\Upsilon}N_{coll}$ of the bottomonium yield in proton-proton (pp) collisions and the number of binary collisions $N_{coll}$. 
Hot medium effects are different for different bottomonium 
states $\Upsilon(1S,2S,3S)$ due to their binding energies. A clear 
sequential suppression pattern has been observed in both p-Pb~\cite{CMS:2022wfi} and Pb-Pb~\cite{CMS:2017ycw,ALICE:2018wzm} collisions in experiments, where bottomonium excited states suffer stronger dissociation in the medium. 

There have been many theoretical models on the market to study the 
heavy quarkonium evolutions in relativistic heavy-ion collisions. The Rate equation model~\cite{Grandchamp:2003uw,Du:2017qkv}, Boltzmann-type transport model~\cite{Zhu:2004nw,Liu:2010ej}, Semi-classical transport model~\cite{Yao:2018sgn,Yao:2020xzw}, and the Statistical hadronization model~\cite{Andronic:2007bi,Andronic:2006ky} consider both quarkonium dissociation and regeneration from the combination of heavy quark and anti-quark in the QGP. In the complex potential model based on Schr\"odinger equation~\cite{Wen:2022utn,Wen:2022yjx,Islam:2020bnp,Islam:2020gdv}, the heavy quarkonium wave function is evolved by taking in-medium complex potentials, which results in the decoherence of bottomonium wave package. 
Open quantum system approaches like the Lindblad equation~\cite{Brambilla:2020qwo} and Stochastic Schrodinger equation~\cite{Akamatsu:2018xim} are also developed to treat heavy quarkonium and quark as an open quantum sub-system in the environment.

In this work, we employ the well-developed Boltzmann transport model to study the bottomonium dynamical evolutions in p-Pb and Pb-Pb collisions consistently.  Hot medium effects such as the color screening effect and the gluon-dissociation are included in the decay rate of heavy quarkonium~\cite{Chen:2018kfo}. The cold nuclear matter effects like the shadowing effect~\cite{Mueller:1985wy} and Cronin effect~\cite{Cronin:1974zm} are also included in the initial conditions of the transport equation. 
The details of the transport model and the hot medium evolutions are introduced in Section II. 
In Section III, theoretical results are compared with the experimental data from p-Pb and Pb-Pb collisions. A final conclusion is given in Section IV.

\section{Transport model}
In the quark-gluon plasma, dynamical evolutions of bottomonium in phase space can be described with the transport model. After including bottomonium dissociation and regeneration in the hot medium, bottomonium distribution $f_\Upsilon$ in phase space satisfies the equation~\cite{Zhou:2014kka}, 
\begin{align}
    \label{transport}
&\left[ \cosh(y-\eta)\partial_\tau + {\sinh(y-\eta)\over \tau}\partial_\eta+{\bm v}_T\cdot \nabla_T \right] f_\Upsilon\nonumber\\
&=  -\alpha_\Upsilon f_\Upsilon +\beta_\Upsilon,
\end{align}
where $y=1/2\ln[(E+p_z)/(E-p_z)]$ and $\eta=1/2\ln[(t+z)/(t-z)]$ are the rapidities in the momentum and coordinate space. $\tau=\sqrt{t^2-z^2}$ is the proper time. The third term ${\bf v}_T\cdot {\bf \bigtriangledown}_Tf_\Upsilon$ represents the diffusion of bottomonium in phase space with a constant transverse velocity ${\bf v}_T$. Hot medium effects dissociate bottomonium with a rate $\alpha_\Upsilon$,
\begin{equation}
\label{lab-decayrate}
\alpha_\Upsilon ={1\over 2E_T} \int {d^3{\bf k}\over {(2\pi)^32E_g}}\sigma_{g\Upsilon}({\bf p},{\bf k},T)4F_{g\Upsilon}({\bf p},{\bf k})f_g({\bf k},T)
\end{equation}
where $E_T=\sqrt{m_\Upsilon^2+p_T^2}$ is the transverse energy of bottomonium, with the mass $m_{\Upsilon(1S,1P,2S,2P,3S)}=(9.46,9.89,10.02,10.25,10.35)$ GeV taken from Particle Data Group~\cite{ParticleDataGroup:2022pth}. 
${\bf k}$ is the momentum of gluon. 
Bottomonium decay rate induced by the gluon-dissociation is proportional to the gluon density $f_g$. It is taken to be a massless Bose distribution. $F_{g\Upsilon}$ is the flux factor in the reaction. The dissociation cross section of bottomonium in vacuum can be calculated via the method of the Operator-Product-Expansion~\cite{Peskin:1979va,Bhanot:1979vb}. The formula of the dissociation cross-section is written as~\cite{Zhu:2004nw}, 
\begin{align}
    \sigma_{g\Upsilon(1S)\rightarrow b\bar b}(w)&= A_0 {(x-1)^{3/2}\over x^5}      
\end{align}
with the definition 
$x\equiv w/\epsilon$. $w=p_\Upsilon^\mu k_{g\mu}/m_\Upsilon$ 
is the energy of gluon in the rest frame of bottomonium moving with a four-momentum $p_\Upsilon^\mu$. $\epsilon$ is the binding energy of $\Upsilon(1S)$. It is defined as $\epsilon(0)=2m_b-m_\Upsilon$ in vacuum and becomes smaller in the hot medium due to the color screening. Bottom quark mass is taken as $m_b=5.28$ GeV. We approximate the in-medium binding energy of $\Upsilon(1S)$ to be 40\% of its vacuum value to consider the mean color screening effect and neglect the temperature dependence. The detailed temperature dependence of the decay rate is encoded in the gluon density $f_g({\bf k},T)$~\cite{Chen:2018kfo}. $A_0=(2^{11}\pi/27)(m_b^3\epsilon_\Upsilon)^{-1/2}$ is a constant factor. 

Bottomonium can also be produced via the combination of bottom and anti-bottom quarks in the QGP, represented by $\beta_\Upsilon$. It is proportional to the densities of bottom and anti-bottom quarks and also their coalescence probability~\cite{Chen:2017duy} which is connected with the dissociation rate with the detailed balance. As the bottom quark yield is smaller compared with the case of charm quarks, we neglect bottomonium regeneration in both p-Pb and Pb-Pb collisions. 

The initial distribution of bottomonium in nucleus-nucleus collisions can be treated as a superposition of effective nucleon-nucleon collisions. The production cross sections of different bottomonium states have been measured in experiments by CMS~\cite{CMS:2010wld,CMS:2013qur}, ATLAS~\cite{ATLAS:2012lmu}, ALICE~\cite{ALICE:2015pgg} and LHCb~\cite{LHCb:2012aa,LHCb:2014dei,LHCb:2014ngh,LHCb:2013itw} Collaborations respectively at the LHC energies. Take the branching ratios of feed-down processes from Particle Data Group~\cite{ParticleDataGroup:2022pth}, one can extract the direct production cross-section of bottomonium before the feed-down process, $d\sigma_{\rm direct}(1S,1P,2S,2P,3S)/dy=(37.97,44.2,18.27,37.68,8.21)$ nb at $\sqrt{s_{NN}}=5.02$ TeV~\cite{Wen:2022yjx}. To fit the momentum distribution of bottomonium measured by the above Collaborations, the normalized momentum distribution of $\Upsilon(1S)$ 
can be parametrized with the formula, 
\begin{equation}
\label{eq:pp-input}
{dN_{pp}^{\Upsilon}\over 2\pi p_T dp_T} = 
{(n-1)\over \pi (n-2) \langle p_T^2\rangle_{pp}} [1+{p_T^2\over (n-2) \langle p_T^2
\rangle_{pp}}]^{-n} 
\end{equation}
where the mean transverse momentum square and $n$ are parametrized to be $\langle p_T^2\rangle_{pp}=80\ \rm{(GeV/c)^2}$ and $n=2.5$~\cite{Wen:2022yjx} in the central rapidity of pp collision at $\sqrt{s_{NN}}=5.02$ TeV. As bottomonium is produced in parton hard scatterings, the initial spatial distribution of $\Upsilon(1S)$ is proportional to the number of nucleon binary collisions $n_{coll}({\bf x}_T)$. In nucleus-nucleus and proton-nucleus collisions, the initial condition of bottomonium is also affected by the cold nuclear matter effect. For example, the Cronin effect can be included via the Gauss smearing method with the replacement of $\langle p_T^2\rangle_{pp}+g_{gN}\langle l\rangle$ in $dN_{pp}^{\Upsilon}/2\pi p_Tdp_T$. $\langle l\rangle$ is the mean path length of partons travelling in the nucleus before the production of $b\bar b$ dipole. $a_{gN}=0.15\ \rm{(GeV/c)^2}$~\cite{Chen:2016dke} is the square of energy a parton obtains per unit length. Meanwhile, 
nuclear parton density is changed by surrounding nucleons. This shadowing effect also affects the production of bottomonium in Pb-Pb and p-Pb collisions. The modification factor from the shadowing effect is calculated from the EPS09 package~\cite{Eskola:2009uj} and is multiplied in the initial distributions of bottomonium before the start of the transport equation~\cite{Chen:2016dke}. The normalized transverse momentum distributions of bottomonium excited states are taken to be the same as the case of the ground state due to their similar masses.

The temperature profiles of the hot medium created in p-Pb and Pb-Pb collisions at $\sqrt{s_{NN}}=5.02$ TeV are given by the (2+1) dimensional ideal hydrodynamic model. In p-Pb collisions, the maximum initial temperatures of the QGP in forward and backward rapidities are extracted to be $T_c({\bf x}_T=0, b=0)=248$ MeV and $289$ MeV respectively in the most central collisions~\cite{Wen:2022utn,Liu:2013via,Du:2018wsj,Zhao:2020wcd}. $\tau_0=0.6$ fm/c is the start time of hydrodynamic equations where the medium is assumed to reach local equilibrium. While in Pb-Pb collisions, the maximum initial temperatures of QGP are extracted to be 510 MeV in the central rapidity~\cite{Zhao:2017yhj}. The critical temperature of phase transition between QGP and hadronic gas is fixed to be $T_c=165$ MeV at the zero baryon chemical potential. From hydrodynamic equations, the lifetimes of QGP in p-Pb and Pb-Pb collisions are around 3 fm/c and 12 fm/c respectively in the central collisions with $b=0$. Therefore, bottomonium suffers stronger suppression in Pb-Pb collisions compared with the case in p-Pb collisions, reflected in their $R_{AA}$s.

\section{Numerical results in p-Pb and Pb-Pb collisions}

\begin{figure}[!htb]
\includegraphics[width=0.37\textwidth]{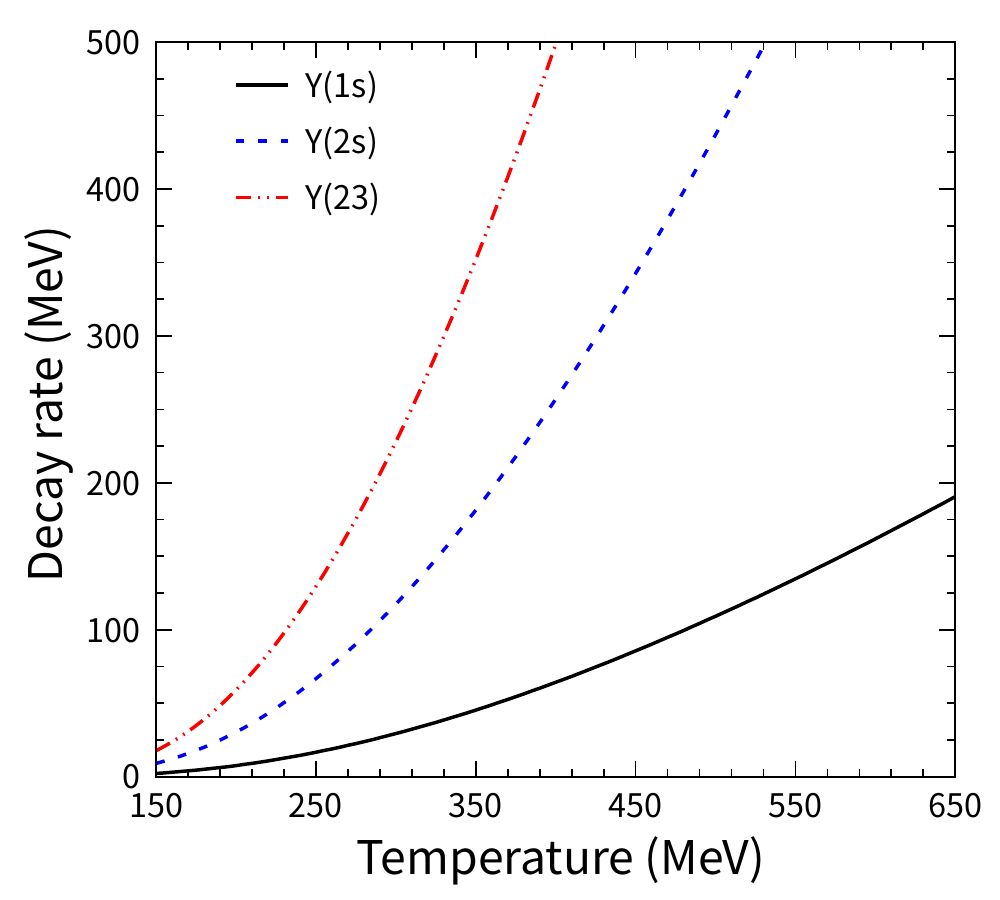}
\caption{ Decay rates of $\Upsilon(1S,2S,3S)$ as a function of temperature in the 
hot medium. The in-medium binding energy of $\Upsilon(1S)$ is taken to be $40\%$ of the vacuum value to consider the color screening effect. The decay rates of the excited states $\Upsilon(2S,3S)$ are obtained by the geometry scale with the ground state. 
}
\label{fig-decay-rates}
\end{figure}

Dynamical evolutions of bottomonium and the QGP have been given by the transport model and hydrodynamic model respectively. In the calculation of bottomonium nuclear modification factors, one of the important ingredients reflecting the interactions between bottomonium and the hot medium is the decay rate of $\Upsilon(1S,2S,3S)$. After considering both color screening and gluon-dissociation processes, the decay rates of bottomonium $\Upsilon(1S,2S,3S)$ are plotted in Fig.\ref{fig-decay-rates}. The decay rate of the ground state is calculated with Eq.(\ref{lab-decayrate}). Decay rates of excited states are obtained via the geometry scale with the ground state. As we neglect the temperature dependence in the binding energy of bottomonium and take an effective in-medium binding energy for $\Upsilon(1S)$, the temperature dependence in the decay rates mainly comes from the density of gluons, which increases with the temperature.

\begin{figure}[!htb]
\includegraphics[width=0.37\textwidth]{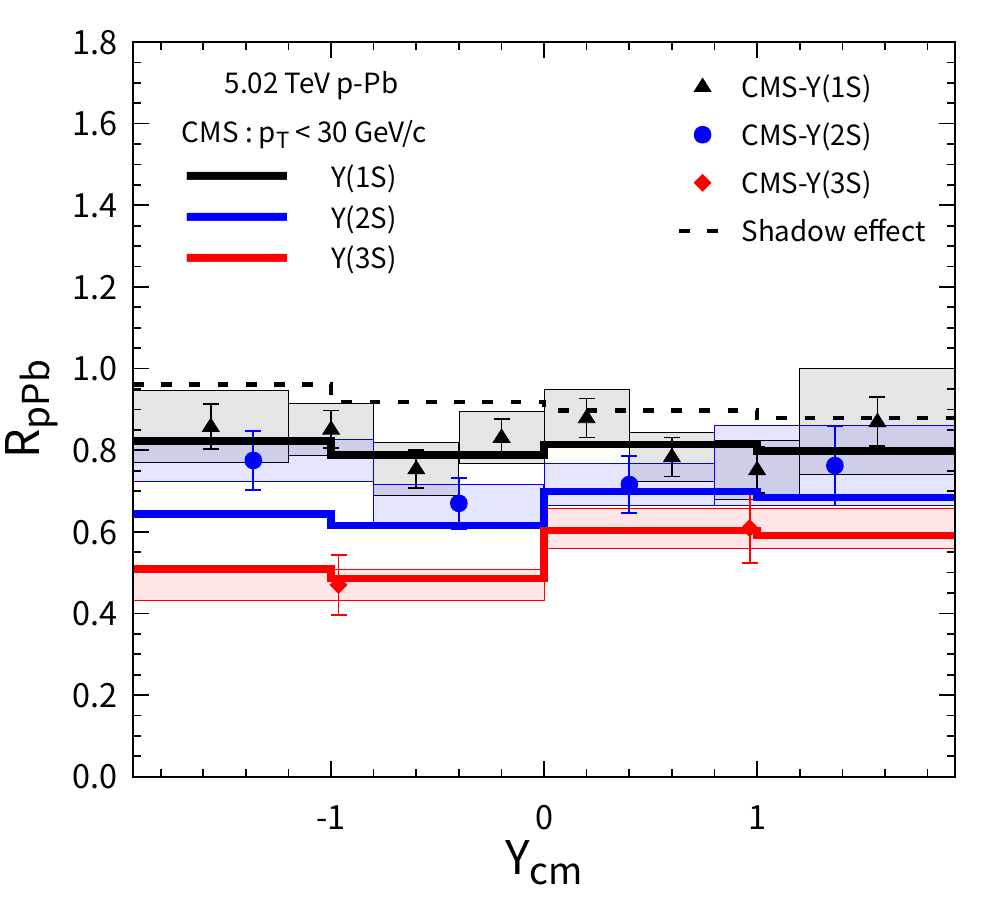}
\caption{ Nuclear modification factors $R_{pPb}$ of $\Upsilon(1S,2S,3S)$ in $\sqrt{s_{NN}}=5.02$ TeV p-Pb collisions. The dashed line includes only cold nuclear matter effects. Solid lines include both cold and hot medium effects. Temperature profiles in forward and backward rapidities are taken in the range $Y_{cm}>0$ and $Y_{cm}<0$ respectively. The experimental data are cited from CMS Collaboration~\cite{CMS:2022wfi}. 
}
\label{lab-pA-Y}
\end{figure}

In p-Pb collisions, after including cold and hot nuclear matter effects, the bottomonium nuclear modification factors are plotted as a function of rapidity in Fig.\ref{lab-pA-Y}. The dashed line only considers cold nuclear matter effects. The shadowing factor becomes different in forward and backward rapidities, which are calculated with the EPS09 package. It is the same for $\Upsilon(1S,2S,3S)$. After considering hot medium dissociation, nuclear modification factors of $\Upsilon(1S,2S,3S)$ become different, plotted with solid lines in the figure. In $Y_{\rm cm}>0$ and $Y_{\rm cm}<0$, two different hydrodynamic profiles are used respectively. In the backward rapidity defined as the Pb-going direction, the anti-shadowing effect enhances the bottomonium production. Meanwhile, medium temperatures in the backward rapidities become higher than the case in the forward rapidities, which gives stronger suppression on bottomonium $R_{pPb}$. The combined effects give similar $R_{pPb}$ of the ground state $\Upsilon(1S)$ in forward and backward rapidities. As excited states are more easily affected by hot medium effects, $\Upsilon(3S)$ $R_{pPb}$ is more suppressed in the backward rapidities. In the pre-equilibrium stage $\tau<\tau_0$, bottomonium also suffers hot medium dissociations where the medium temperatures are approximated to be the value at $\tau=\tau_0$. Feed-down processes from higher P-states and S-states have been included in the calculation.

\begin{figure}[!htb]
\includegraphics[width=0.40\textwidth]{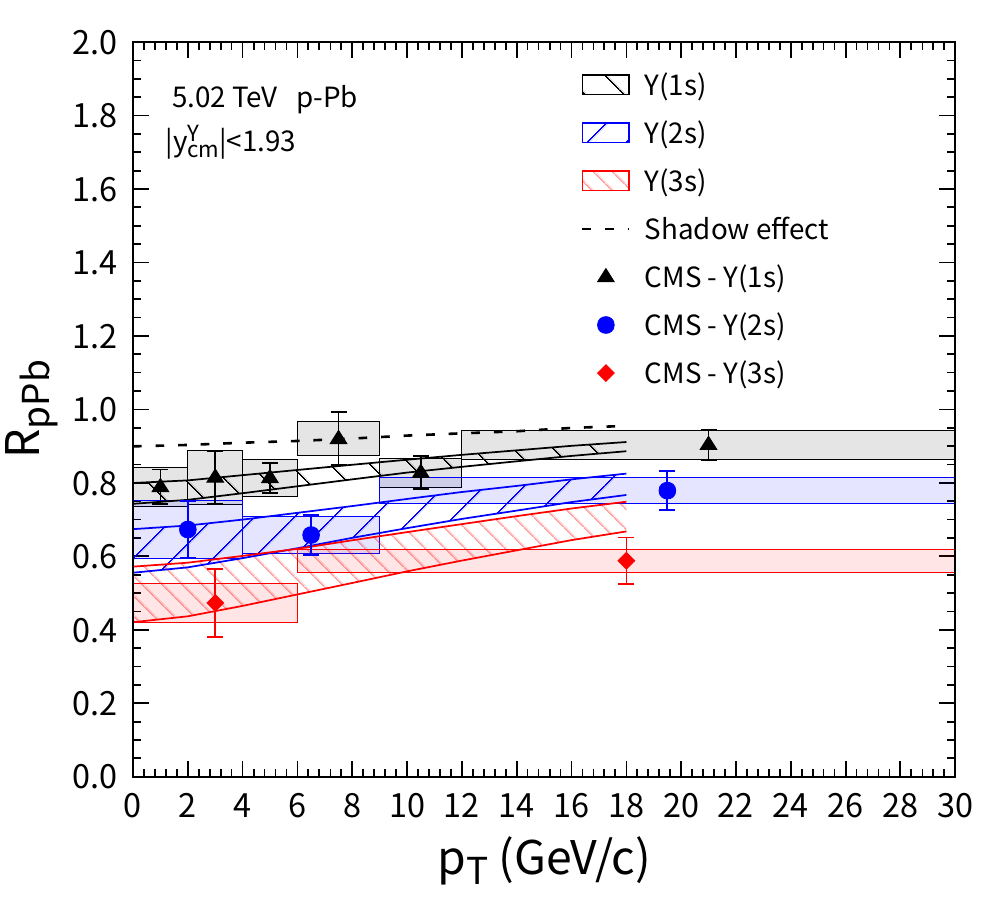}
\includegraphics[width=0.40\textwidth]{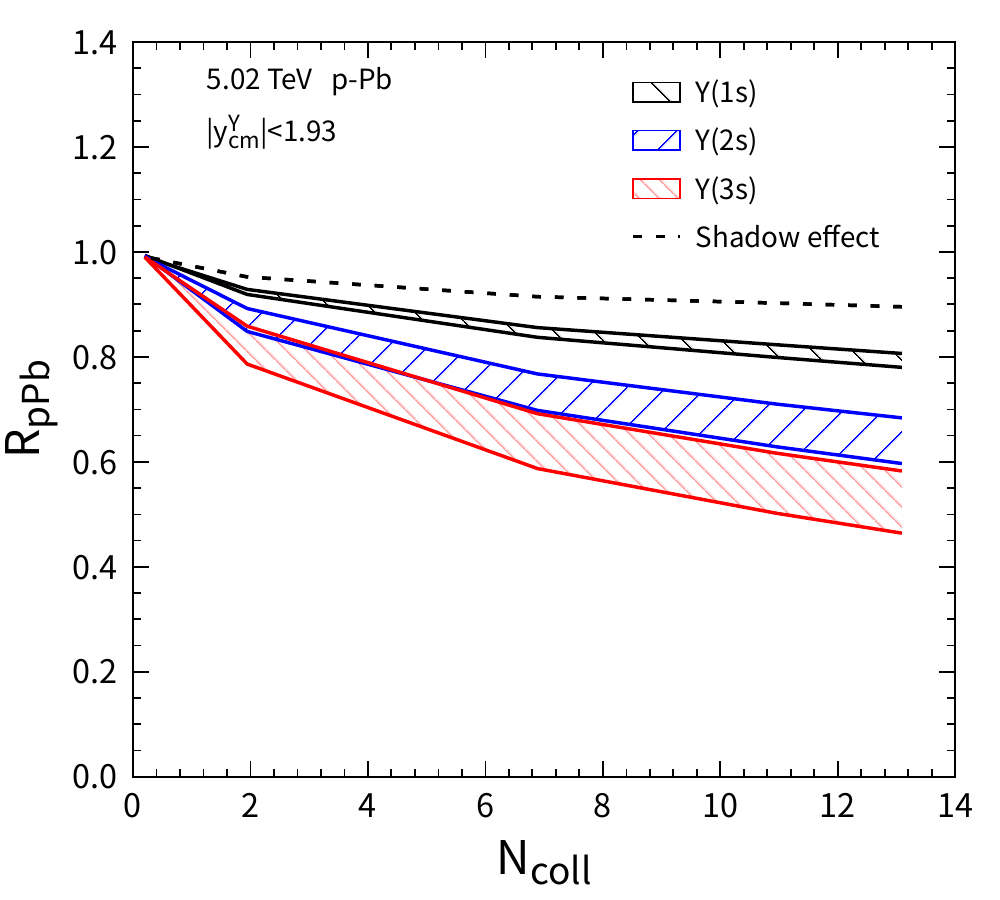}
\caption{(Upper panel) $R_{pPb}$ of $\Upsilon(1S,2S,3S)$ as a function of transverse momentum in $\sqrt{s_{NN}}=5.02$ TeV p-Pb collisions. 
The dashed line includes only cold nuclear matter effects. Theoretical bands correspond to the calculations by taking temperature profiles in forward and backward rapidities of p-Pb collisions respectively. Experimental data are cited from CMS Collaboration~\cite{CMS:2022wfi}. (Lower panel) Bottomonium $R_{pPb}$ as a function of $N_{coll}$. 
}
\label{lab-pA-pt-np}
\end{figure}

The $p_T$ dependence of bottomonium nuclear modification factors is also calculated in Fig.\ref{lab-pA-pt-np}. The dashed line includes only cold nuclear matter effects. As experimental data is in the rapidity range $|y|<1.93$, we use hydrodynamic profiles in forward and backward rapidities respectively. The corresponding results are plotted as the lower and upper limits of the theoretical bands. $R_{pPb}$ increases with the transverse momentum due to the shadowing effect and also the leakage effect where bottomonium with a larger $p_T$ can escape from the hot medium with a shorter time. Hot medium effects result in the difference between $R_{pPb}$s of $\Upsilon(1S,2S,3S)$. Based on the same inputs, we also give the 
calculations about $R_{pPb}$ as a function of $N_{coll}$ in Fig.\ref{lab-pA-pt-np}. The theoretical bands correspond to the calculations with the temperature profiles in forward and backward rapidities.

\begin{figure}[!htb]
\includegraphics[width=0.40\textwidth]{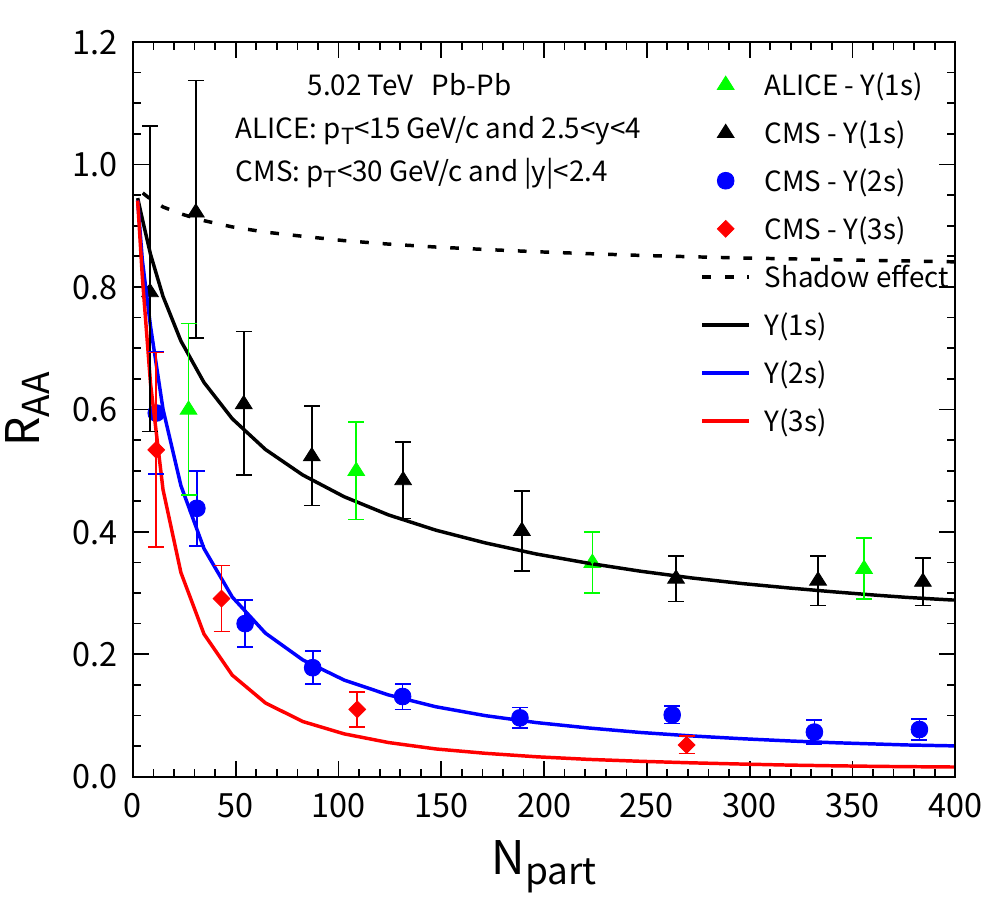}
\includegraphics[width=0.40\textwidth]{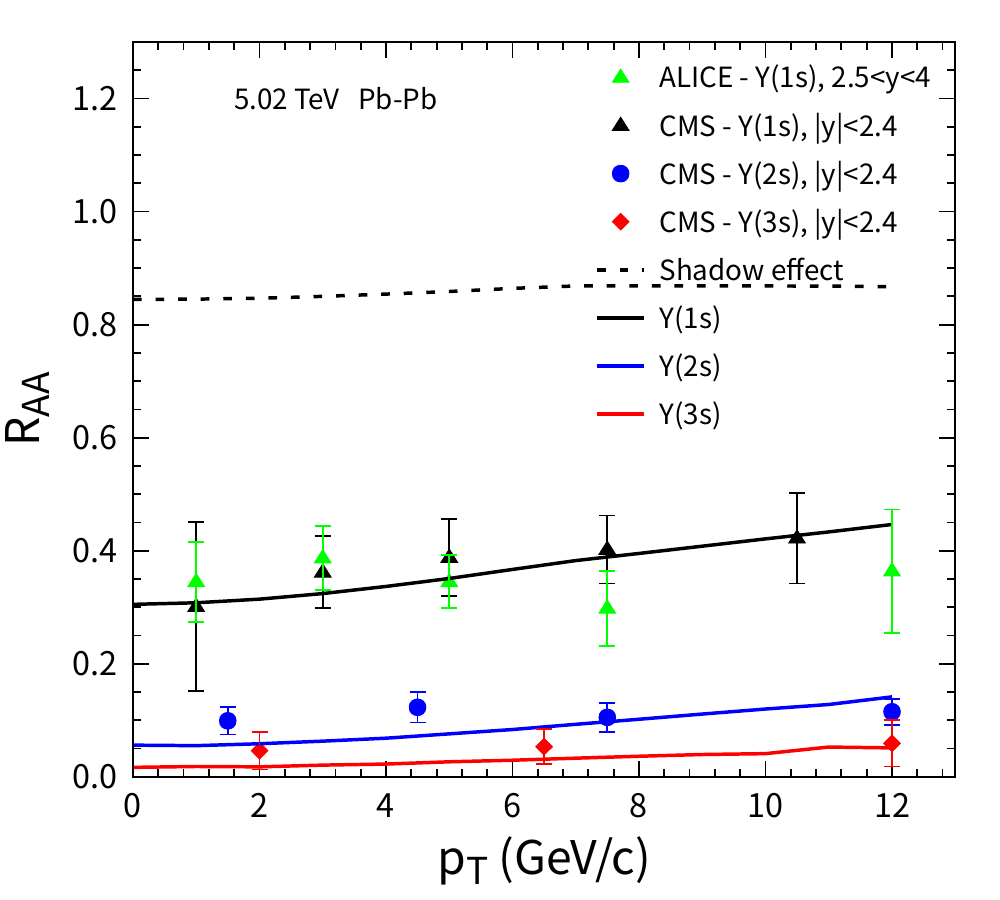}
\caption{  Bottomonium $R_{AA}(1S,2S,3S)$ as a function of $N_p$ (upper panel) and $p_T$ (lower panel) in $\sqrt{s_{NN}}=5.02$ TeV Pb-Pb collisions. Dashed line includes only cold nuclear matter effects. Different solid lines represent the nuclear modification factors of $\Upsilon(1S,2S,3S)$ by taking both cold and hot medium effects and also the feed-down process. The experimental data are cited from CMS~\cite{CMS:2018zza} and ALICE~\cite{ALICE:2018wzm} Collaborations.
}
\label{lab-AA-Np-pt}
\end{figure}

In large collision systems such as Pb-Pb collisions at $\sqrt{s_{NN}}=5.02$ TeV, the initial temperatures of the medium can be as large as $\sim 3T_c$. The density of thermal partons is high sufficient to dissociate most of the bottomonium ground state $\Upsilon(1S)$ and almost all of the excited states in the central collisions. Only those bottomonium states produced at the edge of the fireball can survive the hot medium. We apply the transport model to calculate the bottomonium nuclear modification factors $R_{AA}$ as a function of $N_p$ and $p_T$ in Fig.\ref{lab-AA-Np-pt}. From peripheral to central collisions, hot medium effects on $\Upsilon(1S)$ become stronger. For bottomonium excited states, their $R_{AA}$s are close to zero at $N_p\sim 400$ due to the significant decay rates. There is a slight increase in the $R_{AA}(p_T)$ of $\Upsilon(1S)$, which is due to the leakage effect. In the case without hot medium effects, $R_{AA}$s of $\Upsilon(1S,2S,3S)$ are the same. The evident difference between their $R_{AA}$s indicates a strong hot medium suppression in Pb-Pb collisions. The sequential suppression pattern is clearly observed in both p-Pb and Pb-Pb collisions. 

\section{Summary}
In this work, we employ the Boltzmann transport model to study the nuclear modification factors of bottomonium $\Upsilon(1S,2S,3S)$ at $\sqrt{s_{NN}}=5.02$ TeV p-Pb and Pb-Pb collisions. Cold nuclear matter effects have been included in the initial conditions of the transport model. Hot medium effects such as color screening and gluon-dissociation are considered in the decay rate of the ground state, while the decay rates of excited states are obtained via the geometry scale with the ground state. Theoretical calculations explain well the rapidity and transverse momentum dependence of bottomonium $\Upsilon(1S,2S,3S)$ nuclear modification factors in p-Pb and Pb-Pb collisions, which show a clear pattern of sequential suppression. Those consistent theoretical calculations in different collision systems help to extract reliable decay rates of bottomonium states in the quark-gluon plasma.

\vspace{1cm}
\noindent {\bf Acknowledgement}: 
This work is supported
by the National Natural Science Foundation of China
(NSFC) under Grant Nos. 12175165.

%%%%%%%%%%%%%%%%
%%%%%%%%%%%%%%%%
%%%%%%%%%%%%%%%%%
%%%%%%%%%%%%%%%%

\end{document}